\documentclass{SCIS2024}

\usepackage{tikz}
\usetikzlibrary{arrows,shapes.geometric}
\usepackage{multirow}
\usepackage{amsmath}

\usepackage{CJKutf8}
\AtBeginDocument{\begin{CJK*}{UTF8}{gbsn}}
\AtEndDocument{\end{CJK*}}

\newcommand{\MC}[1]{\textcolor{black}{{#1}}}

\tikzstyle{startstop} = [rectangle, rounded corners, minimum width=3cm, minimum height=1cm, text centered, draw=black]
\tikzstyle{process} = [rectangle, minimum width=3cm, minimum height=1cm, text centered, draw=black]
\tikzstyle{decision} = [diamond, minimum width=3cm, minimum height=1cm, text centered, draw=black]
\tikzstyle{arrow} = [thick,->,>=stealth]

\begin{document}
\ArticleType{RESEARCH PAPER}
\Year{2024}
\Month{}
\Vol{}
\No{}
\DOI{}
\ArtNo{}
\ReceiveDate{}
\ReviseDate{}
\AcceptDate{}
\OnlineDate{}

\title{GPU Acceleration of Numerical Atomic Orbitals-Based Density Functional Theory Algorithms within the ABACUS package}

\author[1,2]{Haochong Zhang}{}
\author[3,4]{Zichao Deng}{}
\author[3]{Yu Liu}{}
\author[3,4]{Tao Liu}{}
\author[3,4]{Mohan Chen}{}
\author[2]{Shi Yin}{{shiyin@iai.ustc.edu.cn}}
\author[1,2]{Lixin He}{{helx@ustc.edu.cn}}

\AuthorMark{Haochong Z}

\AuthorCitation{Haochong Z, Shi Y, Lixin H, et al}



\address[1]{University of Science and Technology of China, Hefei {\rm 230026}, China}
\address[2]{Institute of Artificial Intelligence, Hefei Comprehensive National Science Center, Hefei {\rm 230088}, China}
\address[3]{HEDPS, CAPT, College of Engineering, Peking University, Beijing {\rm 100871}, China}
\address[4]{AI for Science Institute, Beijing {\rm 100080}, China}

\abstract{
\MC{With the fast developments of high-performance computing,}
first-principles \MC{methods based on quantum mechanics} play a significant role in materials research, serving as fundamental tools for \MC{predicting and } analyzing \MC{various} properties of materials. 
\MC{However, the inherent complexity and substantial computational demands of first-principles algorithms, such as density functional theory, limit their use in larger systems.}
The rapid development of heterogeneous computing, particularly General-Purpose Graphics Processing Units (GPGPUs), has heralded new prospects for enhancing the performance and cost-effectiveness of \MC{first-principles algorithms}. 
We utilize GPGPUs to accelerate the \MC{electronic structure } algorithms in Atomic-orbital Based Ab-initio Computation at USTC (ABACUS), a first-principles computational package based on the linear combination of atomic orbitals (LCAO) basis set.
\MC{We design algorithms on GPGPU to efficiently construct and diagonalize the Hamiltonian of a given system, including the related force and stress calculations.}
The effectiveness of this computational acceleration has been demonstrated through calculations on twisted bilayer graphene \MC{with the system size up to} 10,444 atoms.}


\maketitle

\section{Introduction}

Density Functional Theory (DFT)~\cite{64-Hohenberg,65-Kohn} has become one of the most important first-principles methods for understanding and predicting material properties at the atomic scale, as well as for the discovery and design of new functional materials~\cite{3}. 
Recently, with the rapid progress of artificial intelligence (AI) and deep learning, AI for Science (AI4SCI) is increasingly influencing the field of materials science~\cite{ai4sci-1,ai4sci-2,ai4sci-3,ai4sci-4,ai4sci-5,ai4sci-6}. 
%
First-principles computational methods, grounded in fundamental quantum mechanics physical laws and principles, provide reliable data that serve as the foundation for training and validating advanced machine learning models in materials science. 
%
With the increasing demand for data-driven methodologies in materials research, first-principles calculations are poised to maintain their status as a part of computational workflows in the field of materials science. However, the computational time and cost associated with DFT calculations limit their application in solving practical material problems and hinder the accumulation of material data.

The Linear Combination of Atomic Orbitals (LCAO) approach is widely favored in DFT software due to its computational efficiency and its intuitive relationship with molecular orbitals, where electronic wave functions are expressed as linear combinations of atomic orbitals localized around each atom in the system. 
Besides the analytical Gaussian or Slater-type orbitals, several first-principles codes based on numerical atomic orbitals have been developed in recent years, e.g., SIESTA~\cite{siesta}, OpenMX~\cite{OpenMX}, and FHI-aims~\cite{FHI-aims}, to name a few.
One of the principal advantages of LCAO basis sets is their computational efficiency. 
The basis size of atomic orbitals is much smaller compared to other basis sets, such as PW or real-space mesh.
Additionally, atomic orbitals are strictly localized, allowing them to be effectively combined with linear scaling algorithms~\cite{goedecker1999linear} for electronic calculations.
Despite the great computational efficiency of LCAO calculations, further time-consuming challenges arise from matrix diagonalization and the computation of three-center integrals.
While the matrix diagonalization process is well-studied and optimized, the acceleration of three-center integral calculations, particularly through GPU-based approaches, remains a less explored area of research.

Following decades of development, heterogeneous computing systems, particularly General-Purpose Graphics Processing Units (GPGPUs), have emerged as critical tools for enhancing computational efficiency. The exponential growth of data-intensive domains, such as scientific computing \cite{1} and deep learning \cite{2}, has driven the adoption of heterogeneous computing. This collaborative approach allows for the optimal distribution of workloads across Central Processing Units (CPUs) and GPGPUs, improving overall system performance and efficiency. 
%
The acceleration of DFT calculations using Graphics Processing Units (GPUs) has garnered significant attention over the past decade. Early efforts predominantly focused on PW basis sets due to their straightforward implementation on parallel architectures. For instance, the Vienna Ab-initio Simulation Package (VASP)\cite{6,9} integrated GPU support to expedite PW computations, resulting in notable performance enhancements. This GPU acceleration enables calculations that would typically require supercomputing resources to be performed on less powerful computational systems. Similarly, Quantum ESPRESSO\cite{quantum-espresso-1, quantum-espresso-2} utilized GPU acceleration to optimize its PW DFT calculations, achieving substantial speedups in large-scale simulations. The INQ framework~\cite{8}, developed from the ground up with GPU acceleration in mind, has demonstrated the feasibility of employing GPUs for solving the Kohn-Sham equations in both real-time and ground-state DFT applications. In a related effort, Wang et al.~\cite{13,14,15} presented a GPU-accelerated version of the PEtot code designed for large-scale PW pseudopotential calculations on GPU clusters. A parallel implementation of PW DFT has been presented on the new Sunway supercomputer (PWDFT-SW)~\cite{17}. PWDFT-SW achieved a speedup of 64.8x for a physical system containing 4,096 silicon atoms and extended the capabilities of PW-based DFT calculations to large-scale systems containing 16,384 carbon atoms. CP2K~\cite{cp2k} utilizes a mixed Gaussian and PW approach. Researchers have optimized key components of CP2K for GPU architectures, particularly the computation of exact exchange integrals, leading to enhanced performance in hybrid DFT calculations. Sharma et al.~\cite{SPARC} present a GPU-accelerated implementation of the real-space SPARC electronic structure code for performing DFT calculations with the modular math-kernel based implementation for NVIDIA GPUs achieves speedups of up to 6x and 60x in node and core hours respectively compared to CPU-only execution, bringing the time to solution down to less than 30 seconds for a metallic system with over 14,000 electrons. Additionally, the BigDFT project~\cite{big-dft} employed wavelet-based methods compatible with GPU acceleration, providing an alternative approach to Linear Combination of Atomic Orbitals (LCAO) basis sets.

In contrast, fewer studies have addressed GPU acceleration for DFT calculations employing LCAO basis sets. LCAO basis sets are composed of atomic-like orbitals centered on atoms~\cite{lcao-book}. Due to the localized nature of these orbitals, the Hamiltonian and overlap matrices in LCAO methods are often sparse. This sparsity can be exploited to reduce computational costs, especially for large systems, as calculations involve fewer non-zero elements compared to the dense matrices in PW methods. Achieving convergence in LCAO methods often requires a smaller number of basis functions compared to PW methods, which need high-energy cut-offs to capture fine details. This reduction leads to decreased computational resource usage and processing time.
The inherent complexity of LCAO methods, such as the need to handle localized functions and complex overlap integrals, presents unique challenges for parallelization on GPUs. SIESTA~\cite{siesta}, a prominent DFT code using numerical atomic orbitals, has been the subject of efforts to introduce GPU acceleration. Garcia et al.~\cite{siesta} demonstrated improvements in computational efficiency by offloading specific tasks to GPUs, though the full potential of GPU acceleration in this context has not been fully explored. Huhn et al.~\cite{FHI-aims-gpu} present an efficient GPU acceleration strategy for real-space operations in all-electron density functional theory using localized numeric atom-centered basis functions and the domain decomposition method in the FHI-aims code. Focusing on Hamiltonian integration, density updates, and force/stress tensor evaluations, they demonstrate speedups ranging from 2.4x to 6.6x for key steps and 3-4x for full calculations on a 103-material test set, with near-ideal scaling on a 375-atom Bi2Se3 bilayer system.

Despite these advancements, a comprehensive implementation of GPU acceleration specifically tailored for DFT calculations with LCAO basis sets in material simulation is lacking. Existing studies often focus on select aspects or require significant code modifications, limiting their applicability. This paper aims to address this gap by presenting a detailed methodology for integrating GPU acceleration into DFT calculations using LCAO basis sets, optimizing performance while maintaining accuracy and generality.
%
%
In this study, we implement heterogeneous computing algorithms to accelerate DFT computations based on the LCAO basis set in \emph{Atomic-orbital Based Ab-initio Computation at USTC} (ABACUS). ABACUS is an open-source software package based on DFT \cite{4,5}. The package employs both PW and LCAO basis sets in conjunction with norm-conserving pseudopotentials to describe the interactions between atomic nuclei and valence electrons. This work primarily focuses on research and development within the NVIDIA CUDA framework. 
By utilizing these state-of-the-art heterogeneous computing technologies, this paper aims to address performance bottlenecks and computational challenges inherent to first-principles calculations in materials science. 
This reduction in computational cost will enable researchers to address more complex problems that require higher levels of computation while operating within limited resource constraints.

The main contributions of this paper include the following. First, we systematically analyzed the key algorithmic workflows, time complexity, and computational bottlenecks in DFT calculations using the LCAO basis set. Second, a fully optimized framework and method for grid numerical integration GPU acceleration are proposed, significantly improving the computational performance of the critical numerical integration modules. Third, mainstream GPU-supported generalized eigenvalue solver libraries are analyzed and integrated into the system.

The remainder of this paper is as follows. After discussing related works, we begin by providing an overview of the fundamental computational algorithms employed in the LCAO basis set computations within the ABACUS package. This overview is followed by an in-depth analysis of the existing performance bottlenecks that hinder the efficiency of these computations. We then delve into a detailed discussion of the specific computational challenges encountered at different tiers of the ABACUS, introducing the corresponding optimized solutions developed to address these issues. Subsequently, to demonstrate the effectiveness of the implemented optimizations, we present a case study involving twisted bilayer graphene. 
We discuss the computational results obtained from the optimized ABACUS, with a particular focus on the performance improvements achieved through heterogeneous acceleration. The final chapter \ref{Conclusion} summarizes the conclusions of our work while outlining potential future research and developments.

\section{Methods}
\label{Fundamental}

\begin{figure}[!t]
	\centering
\begin{tikzpicture}[node distance=1.8cm,thick,scale=1, every node/.style={scale=0.7}]
\centering

\node (start) [startstop] {Start};
\node (init) [process, below of=start,text width=4cm] {Initialize charge density $\rho_{in}$};
\node (calc1) [process, below of=init,text width=4cm] {Calculate Hamiltonian kinetic term, non-local potential term};
\node (calc2) [process, below of=calc1,text width=4cm] {Calculate Hamiltonian local potential term $V_{\mu \nu}^{\text{loc}}$};
\node (calc3) [process, below of=calc2,text width=4cm] {$H\psi_{i} = \epsilon_i S \psi_{i}$};
\node (calc4) [process, below of=calc3,text width=4cm] {Calculate charge density $\rho_{out}$};
\node (dec1) [decision, diamond, aspect=2, below of=calc4] {$\rho_{out} - \rho_{in} < \rho_{\xi}$};
\node (force) [process, below of=dec1,text width=4cm] {Compute force $F$ and stress $\sigma$ };
\node (dec2) [decision, diamond, aspect=2, below of=force] {$F < F_{\xi}, \sigma < \sigma_{\xi}$};
\node (end) [startstop, below of=dec2] {End};
\node (mix) [process, left of=calc3, xshift=-3cm,text width=4cm] {Mix charge density $\rho_{in}$};
\node (opt) [process, right of=calc4, xshift=3cm,text width=4cm] {Structural optimization};

\draw [arrow] (start) -- (init);
\draw [arrow] (init) -- (calc1);
\draw [arrow] (calc1) -- (calc2);
\draw [arrow] (calc2) -- (calc3);
\draw [arrow] (calc3) -- (calc4);
\draw [arrow] (calc4) -- (dec1);
\draw [arrow] (dec1) -- node[anchor=east] {Yes} (force);
\draw [arrow] (force) -- (dec2);
\draw [arrow] (dec2) -- node[anchor=east] {Yes} (end);
\draw [arrow] (dec1) -| node[anchor=east] {No} (mix);
\draw [arrow] (mix) |- (calc2);
\draw [arrow] (dec2) -| node[anchor=west] {No} (opt);
\draw [arrow] (opt) |- (init);

\end{tikzpicture}
\caption{ABACUS LCAO basis set calculation flow. 
The SCF calculation begins with an initial charge density $\rho_{in}$, which is used to construct the Hamiltonian matrix. The charge density $\rho_{out}$ is obtained by solving the generalized eigenvalue problem through the diagonalization of the Hamiltonian matrix. The mixed charge density is then used to update the Hamiltonian matrix, and the diagonalization is repeated until the charge density converges. Structural optimization consists of multiple SCF iterations, during which the ionic positions and lattice vectors are updated according to the computed forces and stress.}
\label{fig1}
\end{figure}
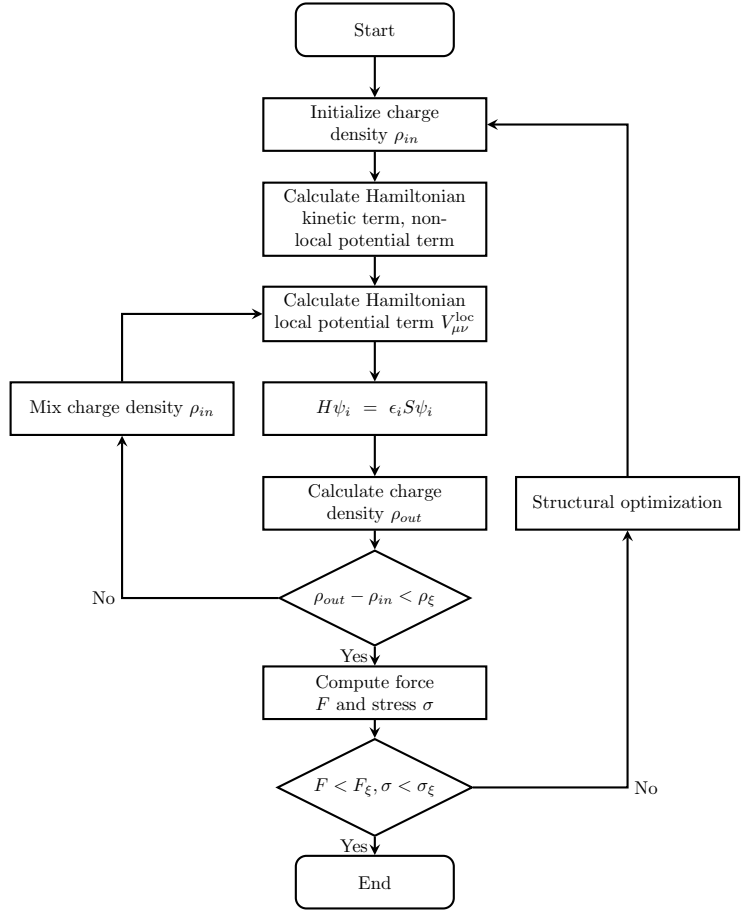

\subsection{Numerical Atomic Orbitals}

ABACUS primarily employs the numerical atomic orbitals to solve the Kohn-Sham equation. As illustrated in Figure \ref{fig1}, the general computational process in ABACUS consists of two main iterative loops: the ionic iteration and the self-consistent field (SCF) iteration.

The ionic iteration, also known as the geometry optimization loop, focuses on finding the equilibrium atomic positions that minimize the total energy of the system. In each ionic iteration, the forces acting on the atoms are computed based on the electronic structure obtained from the SCF iteration. The atomic positions are then updated using optimization algorithms such as the conjugate gradient method or the quasi-Newton method. This process is repeated until the forces on the atoms fall below a specified threshold, indicating that the system has reached a stable geometric configuration. The calculation of forces based on the charge density is a crucial step in the ionic iteration. This process involves performing grid integration in ABACUS, which are computationally intensive.

Nested within each ionic iteration is the SCF iteration, which aims to solve the Kohn-Sham equations self-consistently to obtain the converged charge density and total energy for a given set of atomic positions. The SCF iteration begins with an initial guess of the charge density, which is used to construct the Kohn-Sham Hamiltonian. The Hamiltonian is then diagonalized to obtain the Kohn-Sham eigenvectors and their corresponding eigenvalues. From these eigenvectors, a new charge density is computed and compared with the initial guess. If the difference between the new and old densities exceeds a certain tolerance, the Hamiltonian is updated based on the new density, and the process is repeated until self-consistency is achieved.

The convergence of both the ionic and SCF iterations is crucial for obtaining accurate and reliable results from ABACUS calculations. By iteratively refining the atomic positions and charge density, ABACUS enables researchers to study the structural and electronic properties of materials at the atomic scale, providing valuable insights into their behavior and characteristics. A detailed analysis of the Kohn-Sham equations can be found in \cite{4}. In the computational process described above, the most time-consuming parts are the diagonalization of the Hamiltonian matrix and the three-center integrals used in calculating the local potential term of the Hamiltonian, charge density, and force.

To compute the interactions between atoms in continuous space, LCAO DFT calculations necessitate the use of numerical integration methods. In this study, we simplify the three-center numerical integration process by partitioning the continuous space into uniform grids, effectively transforming the computation from a continuous to a discrete domain. This approach allows for an efficient and streamlined calculation of the three-center integrals, reducing the computational burden associated with this critical step in LCAO DFT calculations.
\subsection{Real-Space Grid Integrals}

The real-space grid integrals play a key role in the LCAO algorithms in ABACUS. Specifically, there are several operations that involve the real-space gridi ntegrals and these procedures take a substantial portion of the total computational time.

First, in ABACUS, to construct the Hamiltonian within the localized basis sets, we need to calculate the local potential term that requires the integral over the entire real space, which takes the form of
\begin{equation}
	V_{\mu \nu}^{\text{loc}} = \int \phi_{\mu} (\mathbf{r}) V^{\text{loc}}(\mathbf{r}) \phi_{\nu} (\mathbf{r}) d \mathbf{r}.
\end{equation}
Here, the local potential $V^{\text{loc}}(\mathbf{r})$ refers to the summation of local pseudopotentials, the Hartree potential and the exchange-correlation potential. The numerical atomic orbitals are $\phi_{\mu}$ and $\phi_{\nu}$.
In practical calculations, to discretize the entire space into grids, the spatial integration becomes a summation over all real-space grids
\begin{equation}
	V_{\mu \nu}^{\text{loc}} = \sum_{\mathbf{r}} \phi_{\mu} (\mathbf{r} - \mathbf{R}_\mu) V^{\text{loc}} (\mathbf{r}) \phi_{\nu} (\mathbf{r} - \mathbf{R}_\nu)dV.
\end{equation}
Here $dV$ refers to the volume of a real space grid in a uniformly discretized real-space grid.

Second, we need to use the real-space integration method to obtain the electronic charge density, which is written as
\begin{equation}
    \rho(\mathbf{r}) = \sum_{\mu\nu} \rho_{\mu \nu}\phi_\mu(\mathbf{r} - \mathbf{R}_\mu) \phi_\nu(\mathbf{r} - \mathbf{R}_\nu) ,
\end{equation}
where $\rho(\mathbf{r})$ represents the electronic charge density at position $\mathbf{r}$, $\phi_i(\mathbf{r} - \mathbf{R}_i)$ is the atomic orbital basis functions centered at atomic positions $\mathbf{R}_i$, $\rho_{ij}$ is the elements of the density matrix, which encapsulate the occupation and overlap of the basis functions.

Third, the Pulay force term due to the basis set dependence on atomic positions also needs real-space integration, which is expressed as
\begin{equation}
F^{L-\text{Pulay}} = -\frac{1}{\Omega} \sum_{R} \sum_{\mu \nu} \rho_{\mu \nu}(\mathbf{R}) \left( \left\langle \frac{\partial \phi_{\mu R}}{\partial \tau_{\nu \mathbf{R}}^{\alpha}} \middle| V^L \middle| \phi_{\nu 0} \right\rangle + \left\langle \phi_{\mu R} \middle| V^L \middle| \frac{\partial \phi_{\nu 0}}{\partial \tau_{\nu 0}^{\alpha}} \right\rangle \right).
\end{equation}
Here, $F^{L\text{-Pulay}}$ denotes the Pulay force, $\Omega$ is the volume of the unit cell, $\rho_{\mu \nu}(\mathbf{R})$ are the density matrix elements in the atomic orbital basis, $\phi_{\mu \mathbf{R}}$ and $\phi_{\nu 0}$ are basis functions at positions $\mathbf{R}$ and the origin, respectively, $\tau_{\nu \mathbf{R}}^{\alpha}$ is the $\alpha$-component of the position vector of atom $\nu$ in cell $\mathbf{R}$, and $V^L$ is the local potential experienced by the electrons. The Pulay force arises because the basis functions depend explicitly on atomic positions.

Fourth, the Pulay stress contribution due to basis set dependence on strain is given by
\begin{equation}
\sigma^{L-\text{Pulay}} = -\frac{1}{\Omega} \sum_{R} \sum_{\mu \nu} \rho_{\mu \nu}(\mathbf{R}) \left( \left\langle \frac{\partial \phi_{\mu R}}{\partial \tau_{\mu R}^{\alpha }} \tau_{\mu R}^{\beta} \middle| V^L \middle| \phi_{\nu 0} \right\rangle + \left\langle \phi_{\mu R} \middle| V^L \middle| \frac{\partial \phi_{\nu 0}}{\partial \tau_{\nu 0}^{\alpha}} \tau_{\nu 0}^{\beta} \right\rangle \right),
\end{equation}
where $\sigma^{L\text{-Pulay}}$ is the Pulay stress tensor, and $\tau_{\mu \mathbf{R}}^{\beta}$ is the $\beta$-component of the position vector of atom $\mu$ in cell $\mathbf{R}$. Similar to the Pulay force, the Pulay stress accounts for the explicit dependence of the basis on atomic positions when calculating the stress tensor, which is essential for studying mechanical properties such as elastic constants and responses to pressure.

In the calculation of the above integrals, we can observe significant differences in the integrands, with variations in both the form and the computational goals. However, we can also identify some structural similarities across these calculations. Firstly, all these integrals are three-center integrals, primarily involving a point in space and the centers of two other atomic orbitals. Secondly, after determining the value of the integrand, the process can often be transformed into a form involving the multiplication of vectors. When processing grids in batches, these vector operations can be converted into matrix operations. Finally, the computational cost of determining the values of the integrands cannot be overlooked, as improper handling of this step may easily become a bottleneck in the overall computation.

\begin{figure}
	\centering
	\includegraphics[width=0.6\linewidth]{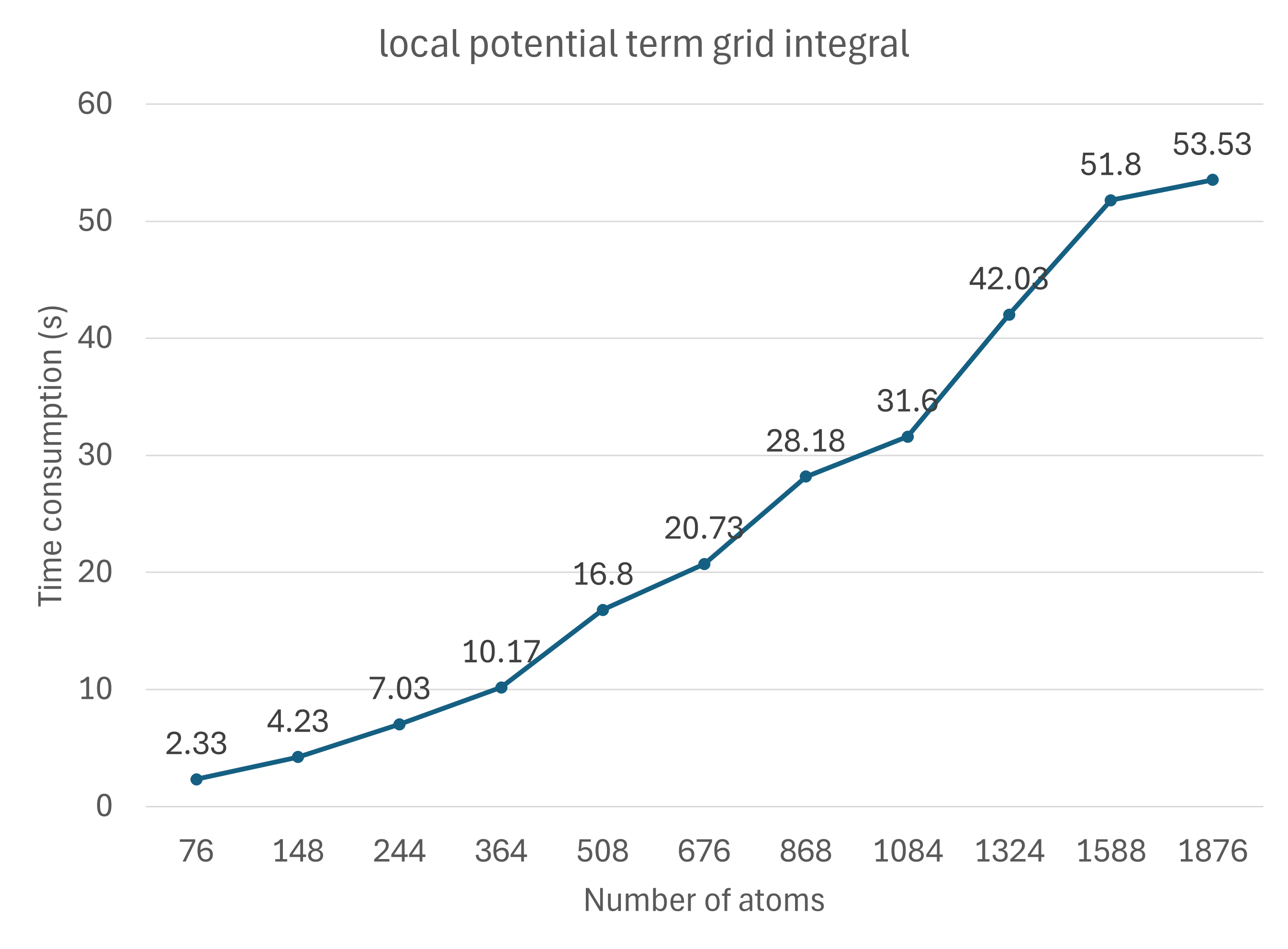}
	\caption{This graph depicts the relationship between the number of atoms and the time consumption for grid integration processes. Starting from 2.41 seconds for 76 atoms, the time consumption rises progressively, reaching 93.95 seconds at 2524 atoms. Since the sparsity of grid integration is fully utilized, the overall grid integration calculation time increases approximately linearly.}
	\label{fig:grid_int_time}
\end{figure}

For a given structure, let $N_g$ denote the number of grids and $N_o$ represent the number of orbitals. In the worst-case scenario, the time complexity of the grid-based integration scales as $O(N_o^2 \times N_g)$. Figure~\ref{fig:grid_int_time} illustrates how the computation time for a single grid integration of the local potential term increases with the number of atoms. As shown in Figure~\ref{fig:grid_int_time}, for common material structures based on the LCAO basis set, the computational process exhibits significant sparsity. This sparsity arises because the cutoff radius of atomic orbitals, relative to the volume of the entire unit cell, is comparatively small in large structures. In such cases, the number of atomic pairs with overlapping cutoff radius typically exhibits a linear growth relationship with the number of atoms.

\begin{figure}
	\centering
	\includegraphics[width=0.6\linewidth]{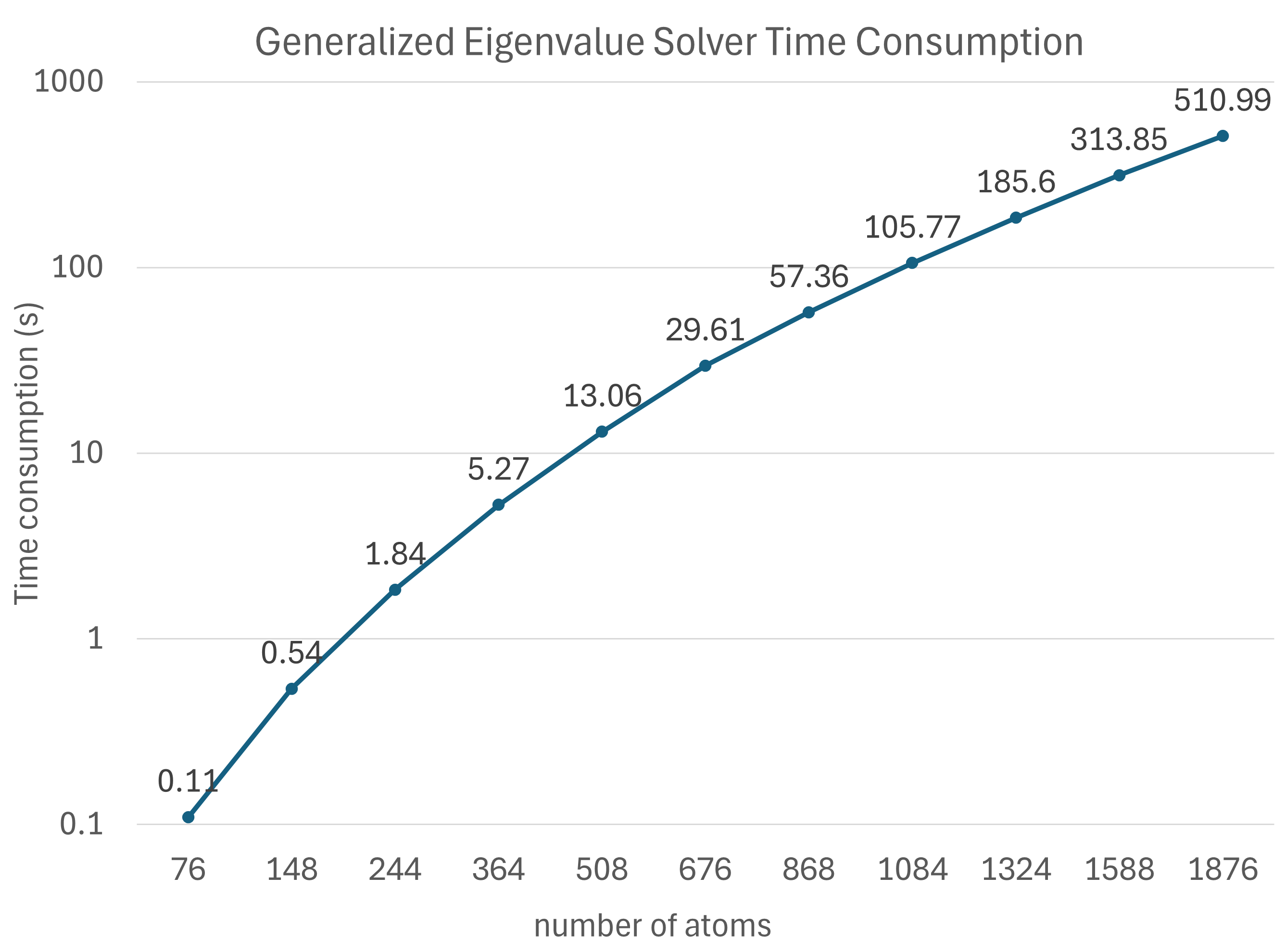}
	\caption{Time consumption of a generalized eigenvalue solver as a function of the number of atoms, plotted on a logarithmic scale. The time required increases steeply with the number of atoms. The logarithmic scale emphasizes the $O(n^3)$ growth in computational time, underscoring the dramatic increase in complexity as the number of atoms grows, particularly for large atomic systems. This reflects the significant computational demand in solving eigenvalue problems as the system size expands.}
	\label{fig:eigen_time}
\end{figure}

The diagonalization of the Hamiltonian matrix is another crucial operation within the SCF iteration, as it yields the Kohn-Sham orbitals and their corresponding eigenvalues. This procedure involves solving the generalized eigenvalue problem:

\begin{equation}
	(H - \varepsilon S)C = 0,
\end{equation}
where \( H \) and \( S \) represent the Hamiltonian and overlap matrices, respectively; \( C \) is the matrix of eigenvectors, and $\varepsilon$ is the diagonal matrix of eigenvalues, defined as
\begin{equation}
	H_{\mu \nu} = \langle \phi_{\mu} | \hat{H} | \phi_{\nu} \rangle, \quad
	S_{\mu \nu} = \langle \phi_{\mu} | \phi_{\nu} \rangle, \quad
	C = (c_{n1}, c_{n2}, \cdots)^{\text{T}}.
\end{equation}
The computational complexity of the diagonalization process exhibits an approximately cubic scaling, \( O(N^3) \), where $N$ represents the size of the matrices involved in the eigenvalue problem. Since $N$ is directly related to the number of atomic orbitals in the LCAO approach, which is positively correlated with the number of atoms, the computational cost increases significantly with system size. Figure~\ref{fig:eigen_time} illustrates how the computation time of the eigenvalue solver increases with the number of atoms. As shown in Figure~\ref{fig:eigen_time}, the computational time required to solve the generalized eigenvalue problem grows steeply with the number of atoms, highlighting the significant computational demand for large systems.

Based on performance tests conducted with the existing ABACUS code employing the LCAO basis set, we evaluated the computational efficiency across various test cases of twisted graphene systems, with sizes ranging from 76 to 1,876 carbon atoms. The CPU tests were performed on a server equipped with two Intel Xeon Silver 4215R CPUs running at 3.20~GHz. All computing cores of both CPUs were utilized via MPI parallelization.

\begin{figure}
	\centering
	\includegraphics[width=1.0\linewidth]{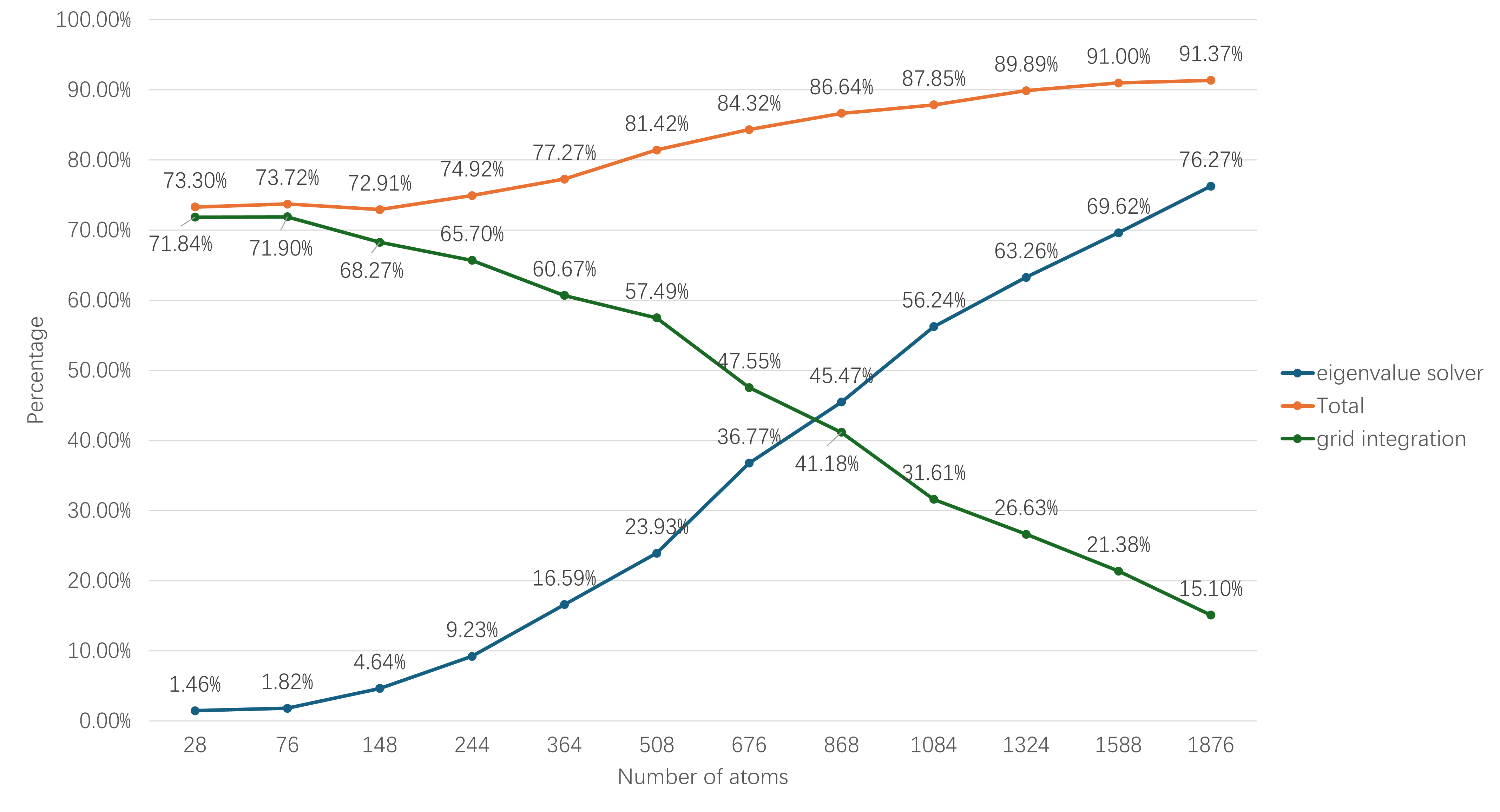}
	\caption{Time ratio distribution for various computational components during a single SCF step, plotted against the number of atoms. The green line represents the time spent on grid integration, which includes the calculation of the local potential term and charge density, in a single SCF step. The blue line represents the time taken by the generalized eigenvalue solver in a single SCF step. The orange line represents the sum of the time for both grid integration and the generalized eigenvalue solver. As the system size increases, there is a marked redistribution in time consumption across components. Initially, the majority of time is spent on grid integration (around 70--75\% for smaller atom counts), but decreases to 15.10\% at 1,876 atoms. Simultaneously, the time proportion for the generalized eigenvalue solver increases from 1.46\% to 76.27\%, reflecting a shift in the computational load as the system grows. }
	\label{fig:cpu-time-ratio}
\end{figure}

Figure~\ref{fig:cpu-time-ratio} shows the change in the time proportion of grid integration and the generalized eigenvalue solver within a single SCF step as the computational scale increases. Specifically, the time proportion for grid integration, which includes the calculation of local potential terms and charge density, decreases from 71.84\% to 15.10\%, while the time proportion for the generalized eigenvalue solver increases from 1.46\% to 76.27\%. The combined time proportion of both components increases from 73.30\% to 91.37\% as the number of atoms in the system grows.

Algorithm analysis and performance test results indicate that for typical material structures, as the system size increases, the proportion of time spent on solving generalized eigenvalues gradually increases. In addition to solving generalized eigenvalues, especially in systems containing hundreds of atoms which is very common in materials calculations, grid integration for the local potential term, charge density, and forces has also become a significant performance bottleneck that cannot be ignored. This highlights the importance of optimizing the eigenvalue solver and grid integration techniques via heterogeneous acceleration to enhance the overall efficiency of the ABACUS package when dealing with larger and more complex material systems.


\subsection{GPU Acceleration of Grid Integrals}

In this work, the optimization of numerical integration over uniform grids using heterogeneous computing focuses on three main aspects: parallel task decomposition of the grid integration problem on multiple computing nodes and multiple GPU chips; synergistic operation between the host and device with balanced workloads; and on-chip performance optimization.

\subsubsection{Task Decomposition on Multi-process}

ABACUS achieves MPI-based multi-process parallelism by partitioning the grid space for efficient grid integration to support multiple computing nodes and multiple GPU chips. For task decomposition in distributed computing, grid integration sections are divided according to the real-space grid, with each process handling approximately the same number of grid computation tasks. Regarding the distribution of computation results, specifically the Hamiltonian matrix elements, ABACUS employs a two-dimensional block-cyclic data layout for matrix element assignments. Since the grid computation results do not directly correspond to the matrix elements, MPI communication is required after  grid integration to exchange individual computation results.

\begin{figure}
	\centering
	\includegraphics[width=0.7\linewidth]{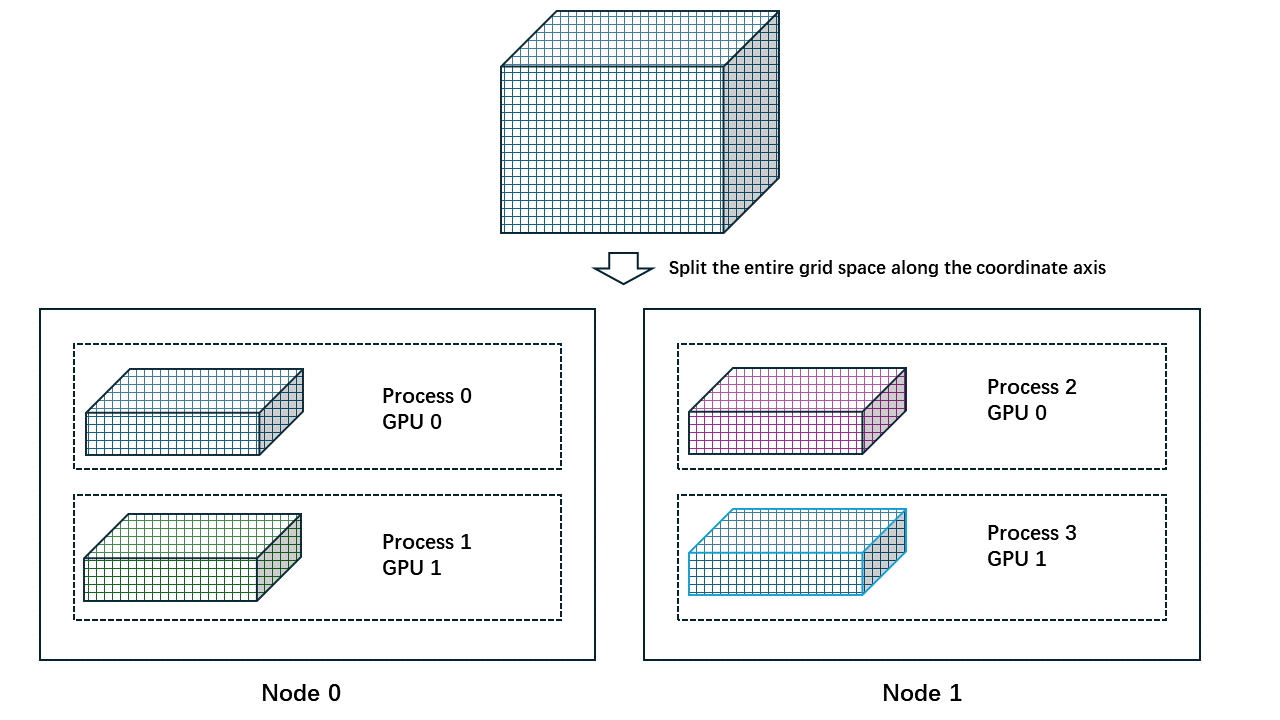}
	\caption{The grid integration task is partitioned and distributed across multiple MPI processes and GPU processors. This diagram assumes that a computational task is assigned to two computing nodes, each utilizing two GPU processors. When decomposing tasks, each process corresponds to a GPU processor, and each process uses OpenMP to implement multi-threaded parallelism.}
	\label{fig:mpipall}
\end{figure}

As shown in Figure \ref{fig:mpipall}, in GPU computing, the number of processes must be greater than or equal to the number of GPU processors. Each process is then bound to a corresponding GPU. A process with an ID of \( P \) is assigned to a GPU with an ID of \( D \), where \( D = P \bmod T \), and \( T \) represents the total number of GPU processors.

\subsubsection{Computational Collaboration Between Host and Device}

\begin{figure}
	\centering
	\includegraphics[width=0.5\linewidth]{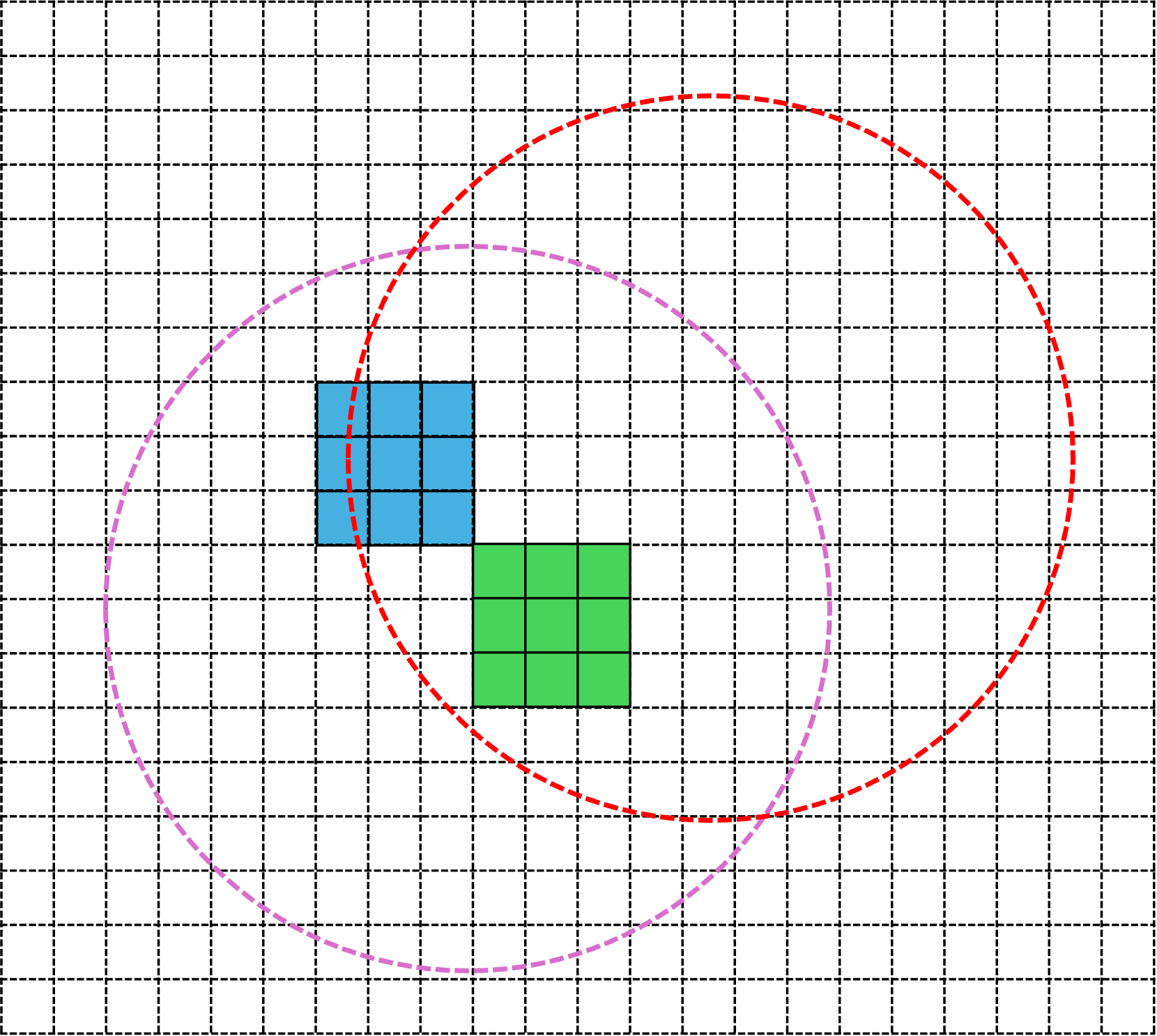}
	\caption{The batch processing method for grid integration within the atomic regions of two atoms whose cutoff radius overlap. Some grids in the blue batch exceed the cutoff radius of one of the atoms, resulting in zero values for the atomic orbitals at these grids. When performing matrix multiplication, the corresponding matrix elements are set to zero. The green batch represents a fully filled matrix because it is entirely contained within the cutoff radius of both atoms.}
	\label{fig:batch_grid}
\end{figure}

By leveraging the sparsity inherent in the LCAO basis set grid integration tasks, we utilize the CPU to decompose complex grid integration into sub-tasks suitable for batch processing. On the GPU, large-scale matrix element computations and batched matrix multiplications are performed, thereby achieving high GPU utilization. CPUs are designed to handle a wide variety of tasks and excel in sequential processing and rapid task switching, while GPUs are optimized for parallel processing, making them ideal for tasks that can exploit their massive core counts for simultaneous calculations. This fundamental difference dictates the kinds of tasks each processor type excels at, influencing how computational workloads are allocated between them.

In the context of numerical atomic orbital basis sets, the influence of each atom on a specific point in space is represented by the contributions of its atomic orbitals at that point. Therefore, for two atoms whose cutoff radii overlap, the grid integration value at a specific grid point can be approximated as the outer product of vectors, where each vector represents the values of the atomic orbitals at that grid point. To enhance computational efficiency, we process grid points in batches, as shown in Figure \ref{fig:batch_grid}. After batch processing, the outer products of multiple vectors transform into matrix multiplication operations between two matrices.

\begin{figure}
	\centering
	\includegraphics[width=1.0\linewidth]{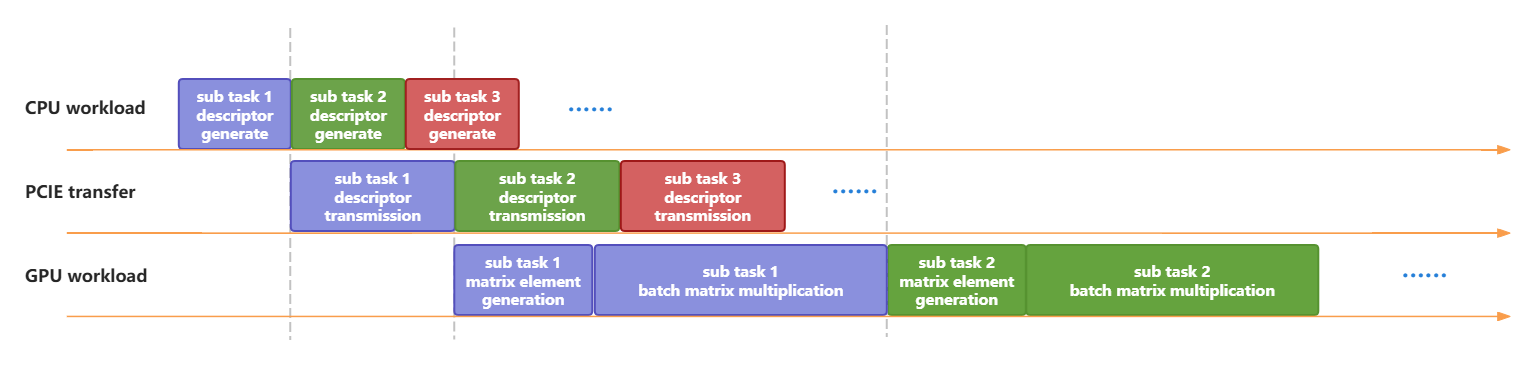}
	\caption{The pipelined parallel execution between the host and device during grid integration. By implementing pipelined parallelism among CPU computation, PCIe data transfer, and GPU computation, the temporal utilization of the GPU has been significantly improved.}
	\label{fig:pipeline}
\end{figure}

The grid integration problem is abstracted into the accumulation of computation results from numerous small matrix multiplications. Complex control computations are handled by the CPU, while the GPU focuses on processing high-throughput computational tasks. As shown in Figure \ref{fig:pipeline}, the entire task consists of a three-stage pipeline executed by the host and device, including the following steps: (1) sub-task division and sub-task descriptor generation on the host; (2) sub-task descriptor transmission to the device via PCIe; and (3) matrix element generation and matrix multiplication computations on the device.

The primary purpose of generating sub-task descriptors is to circumvent complex logical operations on GPUs. These descriptors principally encapsulate critical information required for matrix element computation and matrix multiplication. The computation of matrix elements relies on the relative coordinates of each grid point with respect to atoms and the GPU memory addresses designated for storing these matrix elements. Moreover, batched matrix multiplication necessitates the memory addresses of each small matrix, as well as dimension information such as \( m \), \( n \), and \( k \). To facilitate efficient access during GPU computations, all such descriptor information is sequentially stored in memory according to integer indices. If the distance between a grid point and an atomic nucleus is less than the cutoff radius, the atom is referred to as a 'neighboring atom' for that grid, and the grid is considered a 'neighboring grid' for the atom. Each subtask encompasses a set of grid points in space and their corresponding neighboring atoms.

Apart from the above three-stage pipeline, all computation results are accumulated in the GPU device memory, and once all computational sub-tasks are completed, the results are transferred back to the host memory in a single operation.

\subsubsection{Device Performance Optimization}

The primary computational task on GPU devices involves calculating the values of atomic orbitals at specific grids based on subtask descriptors and subsequently executing matrix multiplication with matrices formed from batched grid values.

Since atomic orbitals are represented using spherical harmonics, calculating the values of atomic orbitals at particular grids primarily involves computing the spherical harmonics based on the relative coordinates between the grids and the atoms. For computing forces and stresses, the derivatives of the spherical harmonics are also calculated. All relevant computational processes have been developed with CUDA kernels to enable GPU acceleration. At runtime, according to CUDA's programming model, fixed grid sizes and block sizes are utilized. Each CUDA thread computes the value of one orbital at one grids in a single calculation, and each thread performs multiple calculations in a loop upon launch. Through the use of subtask descriptors, we ensure a balanced distribution of computational tasks across CUDA threads, with the maximum discrepancy in the number of calculations between different threads not exceeding one spherical harmonic computation. The results of the spherical harmonic computations are stored in GPU memory in a manner optimized for efficient retrieval during matrix multiplication.

The batched matrix multiplication tasks presented here entail specific and unique requirements. For operations within the same batch: (1) There is significant repetition of input matrices. (2) The storage addresses for the output matrices may coincide, leading to memory write conflicts as results accumulate at the same memory addresses. (3) The dimensions of the matrices and the values of alpha may vary. (4) In the vast majority of cases, the dimensions \( m \) and \( n \) of the matrices correspond to the number of atomic orbitals, typically ranging from 1 to 30. The dimension \( k \) represents the number of grids in a grid batch, which varies between 27 and 1000. These unique requirements preclude the direct use of standard batched matrix multiplication libraries, such as cuBLAS, rocBLAS, CUTLASS, and MAGMA~\cite{magma}. However, open-source projects like CUTLASS and MAGMA serve as valuable references for implementing the required functionalities. Specifically, we have adapted and further developed the tiling block strategy from MAGMA's variable-size batched matrix multiplication to meet our specific needs. Input and output matrices are stored as arrays of pointers, which point to the actual memory locations of the matrices, thereby avoiding redundant storage. To prevent computational errors due to write conflicts, atomic operations are employed for accumulating output results. The interface for matrix multiplication has been extended to allow alpha to be an array, enabling individual settings for each matrix. Given the particular matrix sizes involved, it is necessary to finely optimize the tiling strategy to address performance impacts due to changes in matrix size and GPU device memory types.

A dynamic online tile size auto-tuning method is developed. Utilizing template programming, hundreds of CUDA compute kernels with varying tile sizes are automatically generated during the compilation phase based on combinations of tile size dimensions. This approach ensures flexibility during runtime while achieving optimal compile-time optimizations. At runtime, these kernels are executed upon program initialization to dynamically select the kernel with the best performance. Specifically, the program automatically constructs matrix multiplication parameters that are representative of the current task based on user input to test the matrix multiplication kernels. The scale of the test computations remains relatively fixed and does not expand with increases in the computational system size. Practically, the entire auto-tuning process takes no more than two seconds, which is negligible compared to the total execution time of the task.

\subsection{GPU Acceleration of Diagonalization}

After constructing the Hamiltonian matrix of a given system in the numerical atomic orbitals, the next time consuming operation is to diagonalize the matrix and obtian the eigenvalues and eigenfunctions. 
We use GPU to acceerate this operation.
Specifically, mainstream GPU-supported generalized eigenvalue solver libraries have been thoroughly compared, evaluated, and integrated into ABACUS to achieve acceleration.

In ABACUS, the generalized eigenvalue problem takes the following form, where \( \lambda \) represents the eigenvalues and \( X \) represents the eigenvectors. \( H \) and \( S \) are symmetric (Hermitian) \( n \times n \) matrix pairs:
\begin{equation}
	H \times X = \lambda \times S \times X.
\end{equation}
Several mature computational libraries are available for solving the generalized eigenvalue problem. We conduct a thorough and extensive survey of these libraries in our development environment. From a usage perspective, we categorize these libraries into two types: those that support only a single GPU and those that support multiple GPUs. As the size of the input matrices increases, situations may arise where a single GPU's memory is insufficient to accommodate all the computational variables. In such cases, multiple GPU processors are required to accelerate the calculation. However, using multiple GPU processors introduces additional latency due to inter-process communication. This necessitates choosing the appropriate computational approach based on the scale of the problem.

The computational libraries that support only a single GPU include cuSOLVER, rocSOLVER, and hipSOLVER. These libraries are optimized for solving generalized eigenvalue problems on a single GPU device. They utilize the computational capabilities of the GPU to accelerate eigenvalue computations, taking advantage of the GPU's parallel processing power and high memory bandwidth.

Libraries that support multiple GPUs include cuSOLVERMp~\cite{cusolvermp}, ELPA (Eigenvalue SoLvers for Petaflop Applications)~\cite{elpa}, and HPSEPS (High Performance Symmetric Eigenproblem Solvers)~\cite{hpseps}. These libraries are designed to leverage the power of multiple GPUs to solve large-scale generalized eigenvalue problems efficiently. They employ techniques such as matrix partitioning, data distribution, and parallel processing to distribute the workload across multiple GPUs. ELPA and HPSEPS use a CPU+GPU model, where the GPU version offloads certain operations to GPUs. cuSOLVERMp is a pure GPU distributed eigensolver library.

cuSOLVERMp implements an efficient GPU-only parallel divide-and-conquer algorithm to compute eigenvalues and eigenvectors of symmetric tridiagonal systems. ELPA provides both one-stage and two-stage tridiagonal solvers. The two-stage solver is preferred for performance but requires MPI rank oversubscription to GPUs. HPSEPS implements a generalized dense symmetric eigenproblem standardization block algorithm, combining Cholesky decomposition with the traditional standardization algorithm.

For distributed generalized eigenvalue solvers, communication between processes and devices is a major bottleneck in the system. ELPA performs most communication via MPI on the CPUs. Recently, ELPA supports NCCL for NVIDIA GPUs and RCCL for AMD GPUs. cuSOLVERMp uses GPU-aware MPI, NVLink, and NVSHMEM to enable fast GPU-GPU communication without going through the host. cuSOLVERMp is built upon the Communication Abstraction Library (CAL) module, which encapsulates and supports underlying communication libraries such as OpenUCC and NCCL. HPSEPS performs communication via MPI on the CPUs, leveraging the CPU+GPU heterogeneous architecture.

It is worth noting that the aforementioned libraries adhere to the LAPACK (Linear Algebra PACKage) and ScaLAPACK (Scalable Linear Algebra PACKage) interfaces in terms of their APIs. All these libraries support distributed memory parallelism using 2D block-cyclic data distribution of matrices. This adherence facilitates user portability and ease of integration into existing codebases. Researchers and developers familiar with LAPACK and ScaLAPACK can easily adopt these GPU-accelerated libraries without significant modifications to their code.


\begin{table}[!t]
	\footnotesize
	\caption{Comparison of Distributed Generalized Eigenvalue Solvers}
	\label{tab_eigen_solver}
	\centering
\begin{tabular}{|p{3cm}|p{3cm}|p{3cm}|p{3cm}|}
	\hline
	\textbf{Features}                        & \textbf{cuSOLVERMp}              & \textbf{ELPA}                    & \textbf{HPSEPS}                 \\ \hline
	Target Hardware                          & NVIDIA GPUs                      & CPU, Nvidia, AMD and Intel GPUs, Hygon DCUs         & CPU, NVIDIA GPUs and Hygon DCUs \\ \hline
	Data type                                & complex and real in FP32 or FP64 & complex and real in FP32 or FP64 & real in FP64                    \\ \hline
	Natively supported programming languages & C/C++                            & Fortran/C/C++                    & Fortran/C/C++                   \\ \hline
	CPU-GPU affinity                         & One process per GPU              & Multi-process per GPU            & Multi-process per GPU           \\ \hline
	License                                  & Proprietary                      & Open-source (BSD)                & Proprietary                     \\ \hline
\end{tabular}

\end{table}

In summary, ABACUS relies on mature computational libraries to solve the generalized eigenvalue problem efficiently on GPUs. The choice between single-GPU and multi-GPU libraries depends on the scale of the problem and the available GPU resources. Based on the comprehensive performance analysis and comparison, we have chosen to integrate cuSolver, hipSolver, ELPA, and cuSOLVERMp into ABACUS to support generalized eigenvalue solvers on NVIDIA GPUs. Despite the superior diagonalization performance of cuSOLVERMp on NVIDIA GPU clusters~\cite{cusolvermp}, we recommend the adoption of ELPA for large-scale, multi-node, multi-GPU computations. This recommendation is supported by three key considerations regarding performance and usability: (1) In ABACUS, the generalized eigenvalue problem using the LCAO basis does not require solving for all eigenvectors, and ELPA allows users to specify the number of eigenvectors to compute. (2) ABACUS performs self-consistent iterative calculations that require solving eigenvalues and eigenvectors multiple times. ELPA efficiently handles this by avoiding redundant \( S \) matrix decompositions. (3) ELPA provides extensive support for various GPU computing architectures, including those from NVIDIA, AMD, Hygon, and Intel, thereby facilitating adoption across diverse hardware platforms.

\section{\MC{Results and Discussion}}
\label{Experiments}

The acceleration performance of heterogeneous computing in two different environments has been evaluated. We tested the acceleration achieved with a single GPU and examined the performance in a multi-GPU setup. All experiments were performed using twisted bilayer graphene structures of varying sizes as computational benchmarks, focusing on the impact of GPU utilization on the acceleration of SCF iterations and force grid integration.
\subsection{Testing Systems}
\begin{figure}
	\centering
	\includegraphics[width=0.8\linewidth]{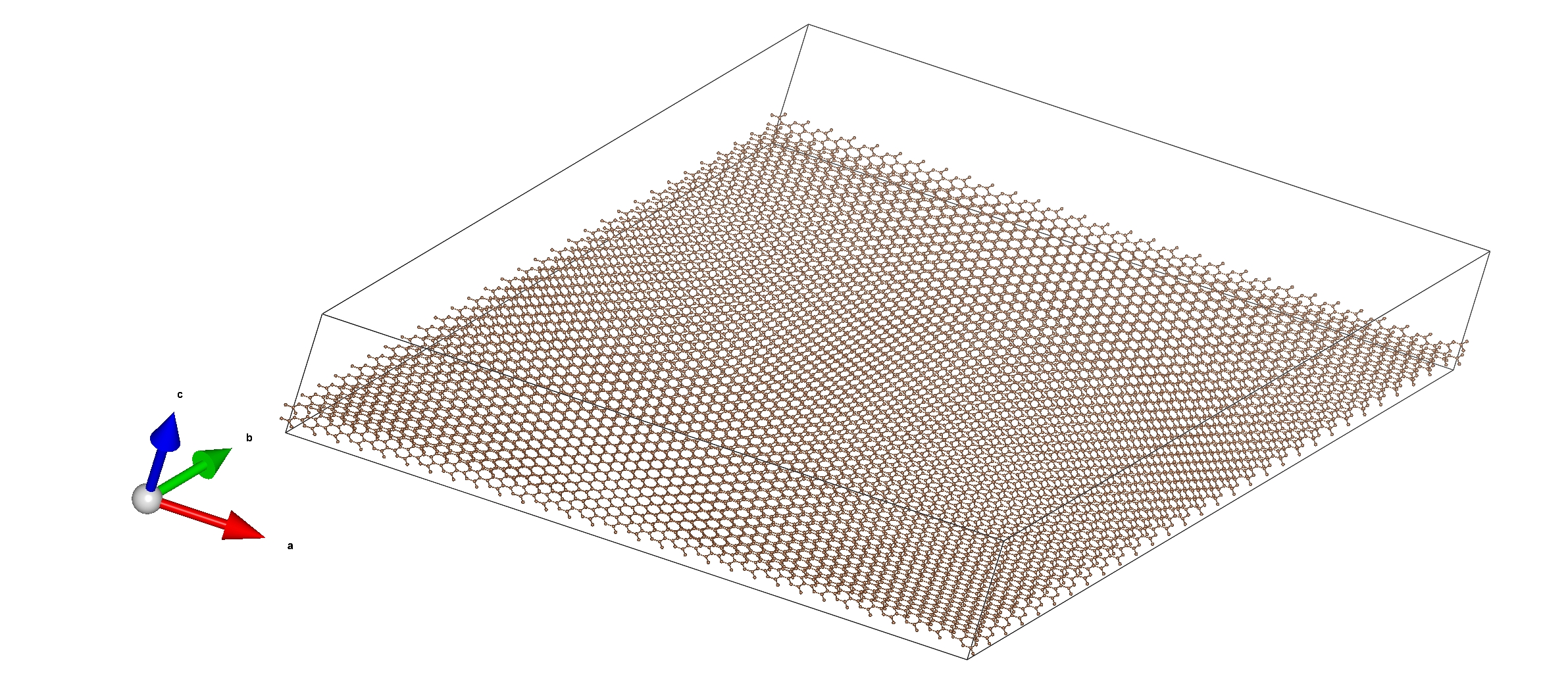}
	\caption{Testing systems: a 10,444 carbon atoms twisted bilayer graphene system}
	\label{fig:10000atoms}
\end{figure}

To validate this work, computational models of twisted graphene with up to 10,444 carbon atoms at various size scales were constructed, as shown in Figure~\ref{fig:10000atoms}. The DFT calculations were performed using a cutoff energy of 100 Rydberg. The convergence criterion for the SCF cycle was set to $1 \times 10^{-6}$. A Gaussian smearing function was applied to the electronic states to enhance convergence, with a smearing width of 0.02 Rydberg, which defines the extent of electronic state broadening. The lattice constant used in the test is 1.8897259886 Bohr. Table~\ref{Lattice_vectors} presents the lattice vectors for several typical systems. The configuration and structure files involved in all tests can be downloaded from the link provided in Chapter~\ref{code-ava}.
\begin{table}[!t]
	\footnotesize
	\caption{Lattice Vectors for Different Structures}
	\label{Lattice_vectors}
	\centering
 \begin{tabular}{|c|c|c|c|}
\hline
\textbf{Atom Number} & \textbf{X Vector} & \textbf{Y Vector} & \textbf{Z Vector} \\ 
\hline
76 & 9.838, 4.260, 0.000 & -11.921, 27.530, 0.000 & 4.919, 2.130, 9.284 \\
\hline
508 & 24.595, 12.780, 0.000 & -13.833, 26.621, 0.000 & 12.298, 6.390, 24.004 \\
\hline
1876 & 46.731, 25.560, 0.000 & -14.396, 26.320, 0.000 & 23.365, 12.780, 46.128 \\
\hline
10444 & 109.448, 61.770, 0.000 & -14.745, 26.126, 0.000 & 54.724, 30.885, 108.838 \\
\hline
\end{tabular}
\end{table}


\subsection{Single GPU Performance}

In a single-machine, single-GPU setup, the size of the GPU memory limits the computational scale. However, for systems up to 1,876 atoms in twisted bilayer graphene, the setup demonstrates a favorable acceleration. The test environment consists of a server equipped with two Intel Xeon Silver 4215R CPUs and an NVIDIA A30 GPU. The code was compiled using compilers and libraries from Intel OneAPI and NVIDIA CUDA.

\begin{figure}
	\centering
	\includegraphics[width=0.8\linewidth]{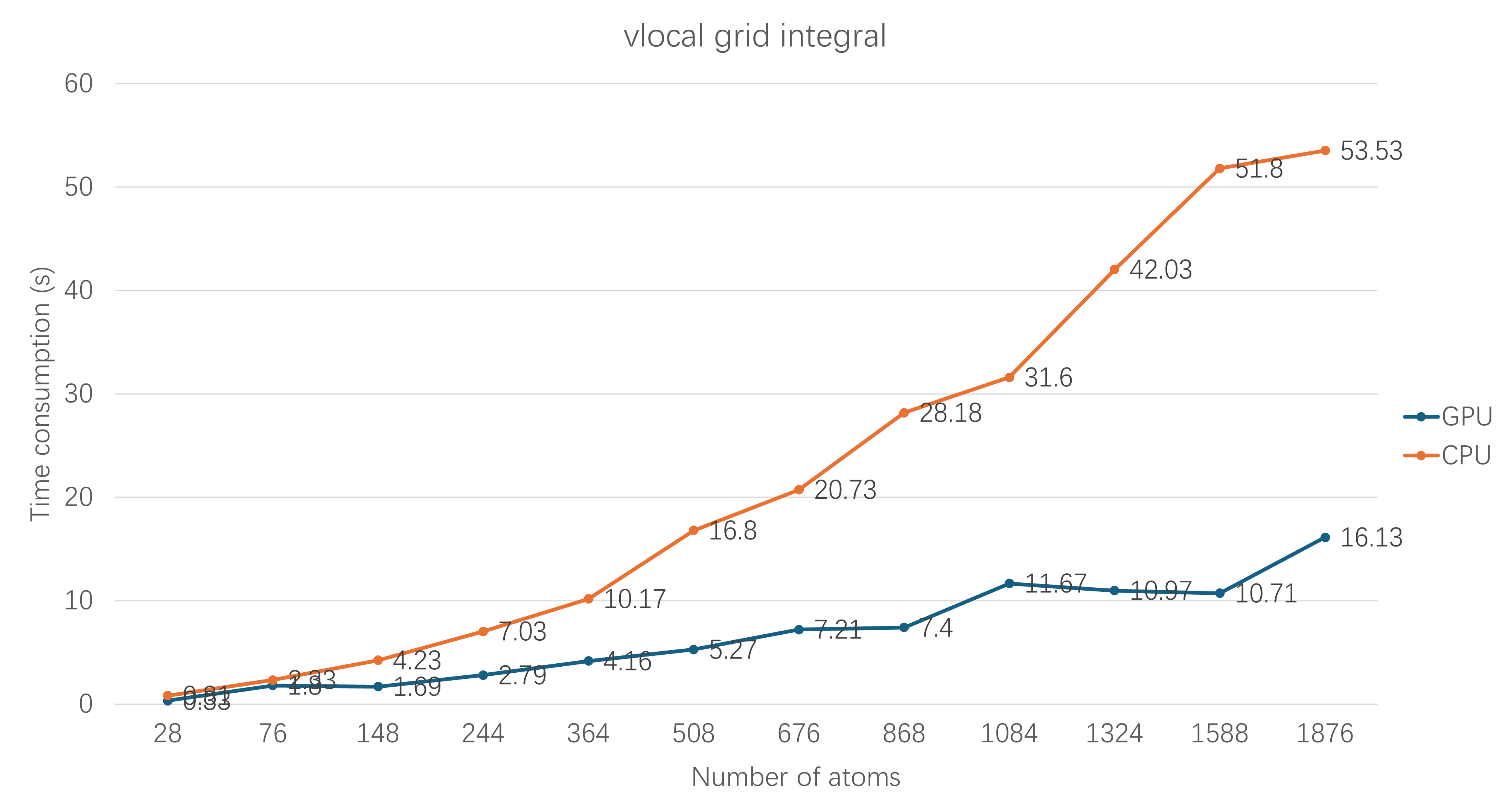}
	\includegraphics[width=0.8\linewidth]{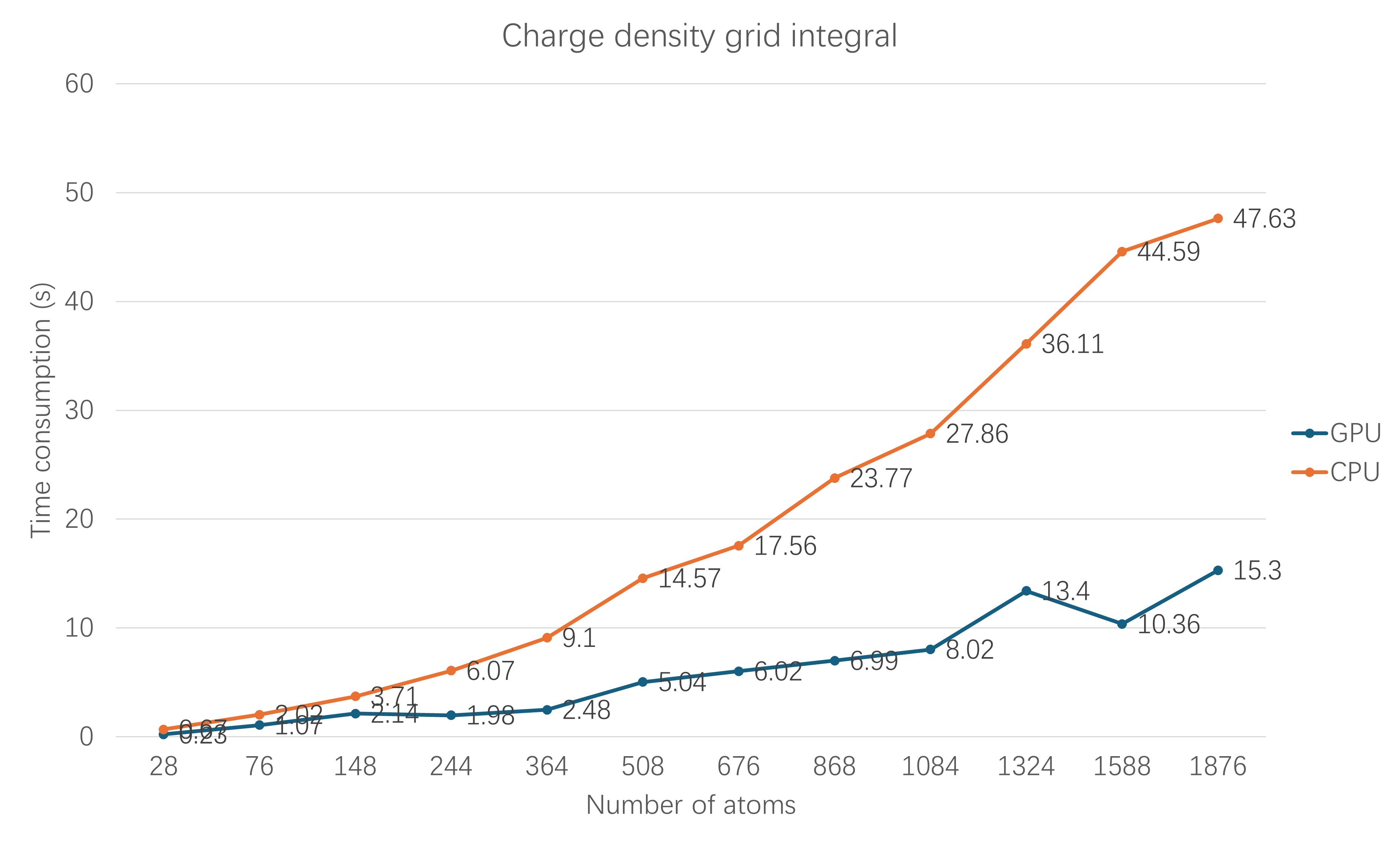}
	\includegraphics[width=0.8\linewidth]{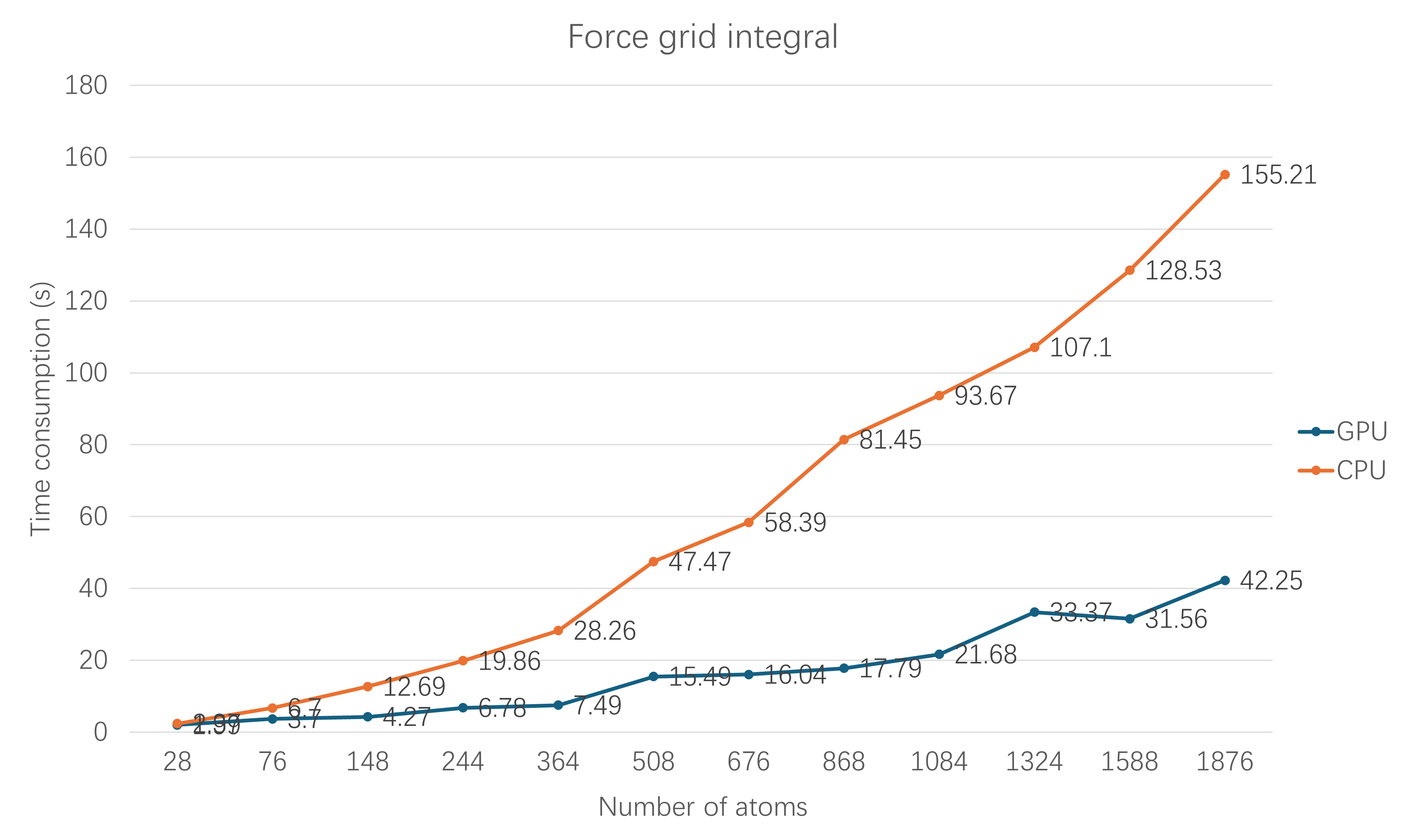}
	\caption{Comparison between an NVIDIA A30 GPU and two Intel Xeon Silver 4215R CPUs for three different types of grid integration: local potential, charge density, and force. The x-axis of all the charts represents the number of atoms, and the y-axis represents the time consumption (in seconds).}
	\label{fig:gpu-vs-cpu}
\end{figure}

\begin{figure}
	\centering
	\includegraphics[width=0.8\linewidth]{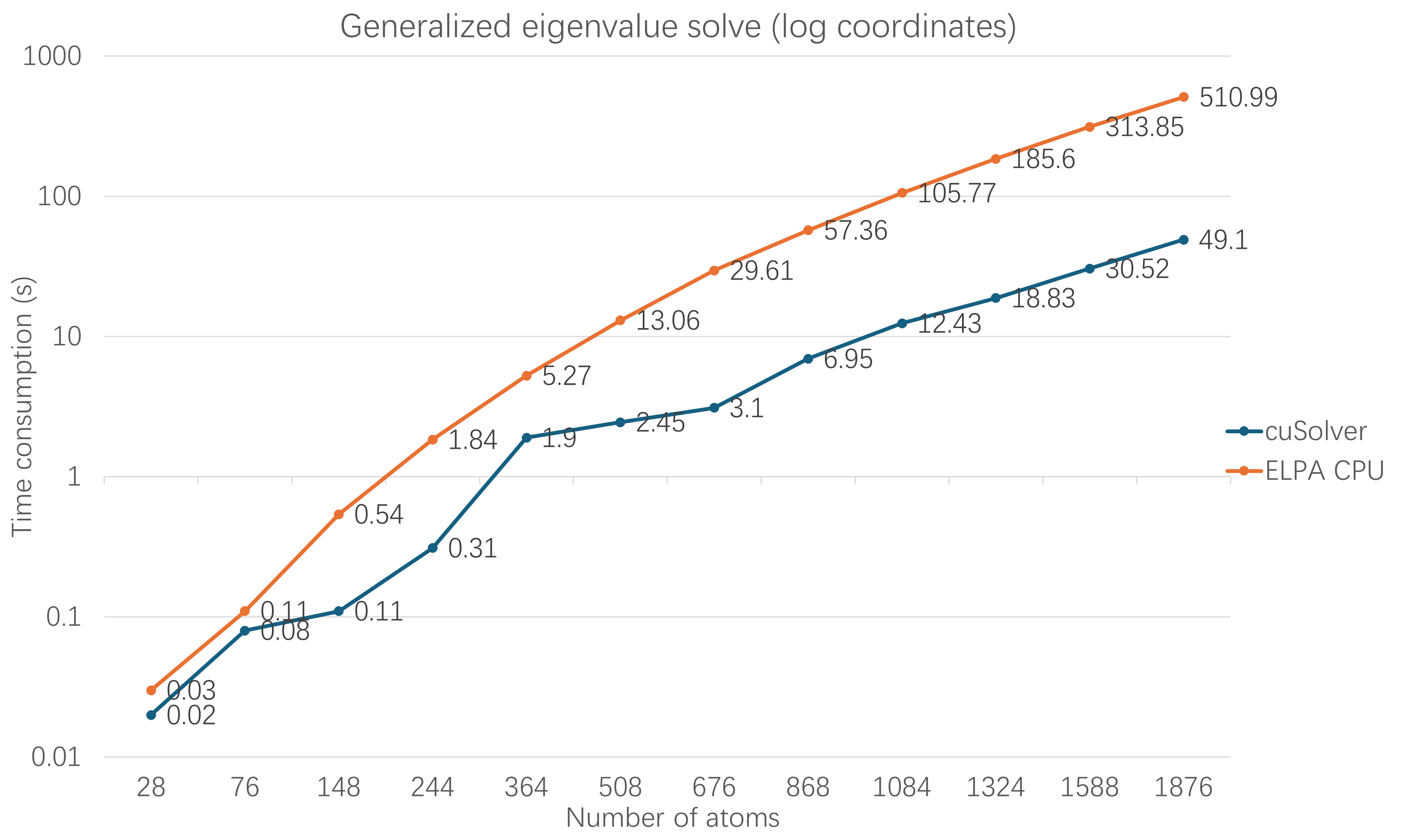}
	\caption{The x-axis represents the number of atoms, and the y-axis (in logarithmic scale) represents the time consumption in seconds. The blue line represents cuSolver with an NVIDIA A30 GPU, and the orange line represents ELPA with two Intel Xeon Silver 4215R CPUs.}
	\label{fig:cusolver-vs-cpu}
\end{figure}

As shown in Figure~\ref{fig:gpu-vs-cpu}, the GPU consistently outperforms the CPU as the number of atoms increases across all grid integration. The benefits of using a GPU become more pronounced as the problem size grows, particularly for larger atomic systems. The charts demonstrate that GPU acceleration scales more efficiently than CPU computation, making it more suitable for handling large-scale atomic simulations. This suggests a clear advantage for using GPUs in computational chemistry or physics simulations involving grid integration. As shown in Figure~\ref{fig:cusolver-vs-cpu}, the GPU method scales much better than the CPU method, with the performance gap widening dramatically as the system size increases. This highlights the advantages of using GPGPUs for computational tasks that involve eigenvalue and eigenvector calculations, particularly for large-scale problems.

To evaluate the effectiveness of automatic tuning in GPU acceleration, we tested the computational performance using different batch sizes while keeping the problem scale fixed. As shown in Table~\ref{tab-diffbatchsize}, due to the different memory tiling strategies selected each time, the variation in the scale of grid batch processing did not lead to significant changes in performance. According to the profiling results obtained using Nsight Compute, the GPU memory bandwidth utilization in the grid integration computation reaches over 97\%, indicating that the grid integration implementation effectively optimizes data movement and maximizes the utilization of the GPU's memory subsystem.

\begin{table}[!t]
	\footnotesize
	\caption{Performance evaluation of grid integration of 1876 atoms in different batch size, time in seconds.}
	\label{tab-diffbatchsize}
	\centering
	\begin{tabular}{|c|c|c|c|}
		\toprule
		Batch size          & Local potential & Charge density & Force \\ \hline
		125  &  53.09    &   46.23     &   154.75     \\
		216  &  52.55    &  49.15    &  158.25    \\
		343  &  54.25    &  49.46    &  157.04    \\
		512  &  55.45    &    48.06    &    162.27 \\ \bottomrule
	\end{tabular}
\end{table}

\subsection{Multi-GPU Performance}

The performance of ABACUS SCF calculations in multi-GPU configurations was evaluated using twisted bilayer graphene samples containing 5,044 and 10,444 carbon atoms as benchmark cases. The test environment consists of servers equipped with two Intel Xeon Scalable 8358 CPUs, 1~TB of DDR4 3200~MHz memory, and eight NVIDIA A100 GPUs (with SXM4 architecture and 80~GB of GPU memory).

First, we evaluated the scalability of various modules within the self-consistent iterative process on multiple GPUs one computation node using a twisted bilayer graphene system containing 5,044 carbon atoms as a test case. As shown in Table~\ref{multi-GPU}, the charge density and local potential grid integration show the most substantial gains from multi-GPU scaling, likely due to their computational complexity and suitability for parallel processing. The ELPA eigenvalue solver sees diminishing returns beyond six GPUs, which suggests either a bottleneck in parallelization or that the task becomes memory-bound or communication-bound at that point.

\begin{table}[!t]
	\footnotesize
	\caption{Performance evaluation of ABACUS using multi-GPU one node configurations (4, 6, and 8 GPU cards), time in seconds.}
	\label{multi-GPU}
	\centering
	\begin{tabular}{|c|c|c|c|}
		\toprule
		Test Module          & 4 GPUs & 6 GPUs & 8 GPUs \\ \hline
		Local potential grid integration &   12.03    &   9.19     &   7.67     \\
		Charge density grid integration       &    9.61    &    5.15    &    4.39 \\
		Eigenvalue solver  &  354.03    &  289.17    &  218.53    \\  \bottomrule
	\end{tabular}
\end{table}

Second, a twisted bilayer graphene system with 10,444 carbon atoms was used to validate the multi-GPU performance across nodes. Experimental data in Table~\ref{10444-atom} show that as the system size expands to tens of thousands of atoms, the difference between the \( \mathcal{O}(n) \) time complexity of grid integration and the \( \mathcal{O}(n^3) \) time complexity of matrix diagonalization becomes increasingly significant. In this context, the eigenvalue solver dominates the computational cost, while the local potential and charge density grid integration account for only a small percentage of the overall computation time.

\begin{table}[!t]
	\footnotesize
	\caption{Performance of 10,444 carbon atoms computed on 3 to 8 nodes equipped with two Intel Xeon Scalable 8358 CPUs and eight NVIDIA A100 GPUs. The value is time in seconds.}
	\label{10444-atom}
	\centering
	\begin{tabular}{|c|c|l|l|l|l|l|}
		\toprule
		               Task                & 24 GPUs & 32 GPUs & 40 GPUs & 48 GPUs & 56 GPUs & 64 GPUs \\ \hline
		One step self-consistent iteration & 943.10  & 749.79  & 664.81  & 588.33  & 519.49  & 457.54  \\
		 Local potential grid integration  &  18.13  & 14.93   & 12.62   & 10.01   & 8.67    & 7.33    \\
		 Charge density grid integration   &  14.54  & 10.66   & 10.32   & 8.72    & 6.33    & 5.57    \\
		        Eigenvalue solver          & 710.67  & 575.30  & 496.34  & 417.38  & 360.98  & 327.13  \\ \bottomrule
	\end{tabular}
\end{table}

To analyze the strong scalability based on the Table~\ref{10444-atom}, we calculate the speedup and efficiency for each task as the number of GPUs increases. The efficiency defined with the speedup and multiple of GPUs:
\begin{equation*}
	\text{Efficiency} = \frac{\text{Speedup}}{\text{N GPUs} / 24}
\end{equation*}

As shown in Table~\ref{strong-one-step}, one step self-consistent iteration shows good scalability, with efficiency remaining above 77\% even at 64 GPUs. The speedup increases consistently with the number of GPUs. As shown in Table~\ref{strong-local-potential} and Table~\ref{strong-charge-density}, local potential grid integration and charge density grid integration exhibit excellent scalability, with efficiency staying above 83\% for all GPU configurations. As shown in Table~\ref{strong-ELPA}, the ELPA eigenvalue solver also demonstrates good scalability, although the efficiency drops to 81\% at 64 GPUs. Overall, the tasks show good scalability up to 8 nodes and 64 GPU cards, with the grid integration tasks scaling particularly well. The ELPA eigenvector solver also scale well, but their efficiency decreases slightly more as the number of GPUs increases. However, due to the increase in inter-process communication, a further decrease in efficiency can be expected as more computing nodes are added.

\begin{table}[!t]
	\footnotesize
	\caption{Strong scalability analysis of one step self-consistent iteration}
	\label{strong-one-step}
	\centering
	\begin{tabular}{|c|c|c|c|}
		\toprule
		GPUs & Time(s) & Speedup & Efficiency \\
		\midrule
		24 & 943.10 & 1.00 & 100\% \\
		32 & 749.79 & 1.26 & 94\% \\
		40 & 664.81 & 1.42 & 85\% \\
		48 & 588.33 & 1.60 & 80\% \\
		56 & 519.49 & 1.82 & 78\% \\
		64 & 457.54 & 2.06 & 77\% \\
		\bottomrule
	\end{tabular}
\end{table}
\begin{table}[!t]
	\footnotesize
	\caption{Strong scalability analysis of local potential grid integration}
	\label{strong-local-potential}
	\centering
\begin{tabular}{|c|c|c|c|}
	\toprule
	GPUs & Time(s) & Speedup & Efficiency \\
	\midrule
	24 & 18.13 & 1.00 & 100\% \\
	32 & 14.93 & 1.21 & 91\% \\
	40 & 12.62 & 1.44 & 86\% \\
	48 & 10.01 & 1.81 & 90\% \\
	56 & 8.67 & 2.09 & 89\% \\
	64 & 7.33 & 2.47 & 93\% \\
	\bottomrule
\end{tabular}
\end{table}
\begin{table}[!t]
	\footnotesize
	\caption{Strong scalability analysis of charge density grid integration}
	\label{strong-charge-density}
	\centering
\begin{tabular}{|c|c|c|c|}
	\toprule
	GPUs & Time(s) & Speedup & Efficiency \\
	\midrule
	24 & 14.54 & 1.00 & 100\% \\
	32 & 10.66 & 1.36 & 102\% \\
	40 & 10.32 & 1.41 & 84\% \\
	48 & 8.72 & 1.67 & 83\% \\
	56 & 6.33 & 2.30 & 98\% \\
	64 & 5.57 & 2.61 & 98\% \\
	\bottomrule
\end{tabular}
\end{table}

\begin{table}[!t]
	\footnotesize
	\caption{Strong scalability analysis of eigenvector solver (ELPA)}
	\label{strong-ELPA}
	\centering
\begin{tabular}{|c|c|c|c|}
	\toprule
	GPUs & Time(s) & Speedup & Efficiency \\
	\midrule
	24 & 710.67 & 1.00 & 100\% \\
	32 & 575.30 & 1.24 & 93\% \\
	40 & 496.34 & 1.43 & 86\% \\
	48 & 417.38 & 1.70 & 85\% \\
	56 & 360.98 & 1.97 & 84\% \\
	64 & 327.13 & 2.17 & 81\% \\
	\bottomrule
\end{tabular}
\end{table}

\section{\MC{Conclusions}}
\label{Conclusion}

This research demonstrates significant advancements in accelerating first-principles calculations using GPGPUs, particularly within the ABACUS framework, which is based on the LCAO basis set. The experiments conducted on various test cases, including large-scale twisted bilayer graphene systems, have validated the effectiveness of GPU acceleration in improving computational efficiency. The results indicate substantial performance improvements, particularly in grid integration and generalized eigenvalue solving, critical components of the ABACUS framework. These optimizations have enabled more efficient resource use and reduced computational costs, facilitating the study of increasingly complex material systems.

In the future, several areas for further development can enhance both the performance and flexibility of the ABACUS software. First, exploring mixed-precision techniques could yield further performance gains, as these approaches leverage the computational speed of lower precision while maintaining the accuracy needed for scientific calculations. 
Second, future efforts could focus on extending grid integration techniques to support more complex and non-uniform grid partitioning schemes, which may improve efficiency in systems with irregular geometries.
Third, extending support for a broader range of hardware, including alternative GPU architectures and custom accelerators, will mitigate risks related to supply chain uncertainties and enable the software to run on diverse computing platforms. 
Fourth, further research should be directed toward designing algorithms that fully exploit advanced hardware features such as Tensor Cores. These optimizations can unlock even greater performance improvements, especially for large-scale simulations.

By addressing these areas, future developments will continue to push the boundaries of computational capabilities in materials science, enabling even larger and more complex systems to be simulated with greater precision and efficiency.

\section{Code Availability}
\label{code-ava}
The source code is available at \url{https://github.com/abacusmodeling/abacus-develop}. The twisted bilayer graphene structures involved in the experiment are available at \url{https://github.com/goodchong/abacus_twist_graphene}.

\Acknowledgements{This work was supported by the Strategic Priority Research Program of the Chinese Academy of Sciences under Grant No. XDB0500201, and by the National Natural Science Foundation of China under Grant Nos. 12134012. The numerical calculations in this paper have been done on the supercomputing system in the Supercomputing Center of University of Science and Technology of China.}

\bibliographystyle{unsrt} 
\bibliography{ref}



\end{document}